\documentclass[11pt]{article}
\usepackage{amsmath,amssymb,color,graphics,epsfig,cite}

\usepackage{amsfonts}

\newcommand{\hoch}[1]{$\, ^{#1}$}

\newcommand{\be}{\begin{equation}}
\newcommand{\ee}{\end{equation}}
\newcommand{\bea}{\setlength\arraycolsep{2pt} \begin{eqnarray}}
\newcommand{\eea}{\end{eqnarray}}
\newcommand{\nn}{\nonumber}

\def\ft#1#2{{\textstyle{\frac{\scriptstyle #1}{\scriptstyle #2} } }}
\def\fft#1#2{{\frac{#1}{#2}}}

\def\0{{\sst{(0)}}}
\def\1{{\sst{(1)}}}
\def\2{{\sst{(2)}}}
\def\3{{\sst{(3)}}}
\def\4{{\sst{(4)}}}
\def\5{{\sst{(5)}}}
\def\6{{\sst{(6)}}}
\def\7{{\sst{(7)}}}
\def\8{{\sst{(8)}}}
\def\sst#1{{\scriptscriptstyle #1}}

\begin{document}

\begin{center}
{\Large {\bf Taub-NUT Black Hole in Higher Derivative Gravity and Its  Thermodynamics } }

\vspace{20pt}

{\large Yu-Qi Chen\hoch{1}, Hai-Shan Liu\hoch{1} and H. L\"u\hoch{1\,,2}}

\vspace{10pt}

{\it \hoch{1}Center for Joint Quantum Studies and Department of Physics,\\
School of Science, Tianjin University, Tianjin 300350, China }

\vspace{5pt}

{\it \hoch{2}Joint School of National University of Singapore and Tianjin University,\\
International Campus of Tianjin University, Binhai New City, Fuzhou 350207, China}

\vspace{40pt}

\underline{ABSTRACT}

\end{center}

We construct the first-order perturbative Taub-NUT black hole solutions in Einstein gravity extended with a cubic curvature invariant. The corrected thermodynamic quantities are then obtained by the standard method and the first law and Smarr relation are satisfied. We also study the perturbative correction to thermodynamics using the Reall-Santos (RS) method and verify that the method is still applicable even though the metrics are no longer asymptotic to Minkowski spacetime. We then apply the RS method to obtain the leading correction to the thermodynamics of the complicated Kerr-Taub-NUT black holes.

\vfill{\footnotesize yuqi\_chen@tju.edu.cn ~~~ hsliu.zju@gmail.com~~~~~mrhonglu@gmail.com }

\thispagestyle{empty}
\pagebreak



\section{Introduction}

The Taub-NUT metric \cite{Taub:1950ez,Newman:1963yy} is one of the most notable exact solutions in General Relativity. It is a simple generalization of the Schwarzschild black hole: in addition to the mass parameter $m$, there is a NUT parameter $n$ that measures the strength of the time bundle over the foliating round $S^2$, with the $SU(2)$ isometry group preserved. The introduction of NUT parameter $n$ has the effect that the metric is no longer asymptotic to Minkowski spacetime, since the time bundle does not degenerate at the asymptotic region. The metric is still asymptotic to local Minkowski spacetime in that the curvature tensors fall off fast enough in the asymptotic region. The NUT parameter brings an advantage that the metric becomes absent from a local curvature singularity as the radial coordinate runs from positive infinity to negative infinity. The price is that the time bundle creates a Misner string singularity \cite{Misner:1963fr}. This singularity can be avoided by imposing a period condition on the time coordinate. However this is not necessary and geodesic can still be completed even in the presence of the Misner string \cite{Clement:2015cxa,Clement:2015aka}. This is because the Misner string, as in the case of the Dirac string, is a global singularity and cannot be observed by local geodesic motions.

Nevertheless, owing to the poor understanding of the Misner string, Taub-NUT metric was traditionally studied in the Euclidean signature, as part of the gravitational instanton that was important to understanding quantum gravity \cite{Gibbons:1979xm}. However, recent works showed that the collapsed objects GRO J1655-40 and M87$\star$ might contain NUT charges \cite{Chakraborty:2017nfu,
Chakraborty:2019rna,Ghasemi-Nodehi:2021ipd}. As suggested in \cite{Chakraborty:2022ltc,Chakraborty:2023wpz}, primordial black holes, as a potential dark matter candidate, could also contain NUT. These results indicate that we should treat the Taub-NUT metric in Lorentz signature more seriously. In Lorentz signature, the Taub-NUT has an event horizon regardless the value of the parameter $m$, be it positive, negative or zero, with the topology of $\mathbb R^2\ltimes  S^2$.

Treating Taub-NUT as a black hole faces a challenge to fit the well established black hole thermodynamics. Requiring that the number of independent thermodynamic variables be the same as the number of parameters, one expects that the first law takes the form
\begin{equation}
    \delta M=T\delta S+\Phi_{n}\delta Q_{n}.\label{firstlaw}
\end{equation}
However, in this differential equation equation, only $(T,S)$ are well established with no ambiguity. The mass $M$, NUT charge $Q_n$ and its thermodynamic conjugate $\Phi_n$ are not obvious to determine. Since the metric is not asymptotic to Minkowski spacetime, the AdM formalism of determining the mass no longer applies. A seemingly natural proposal of treating $m$ as the mass would allow the black hole to have arbitrarily negative mass. There have been several attempts of defining NUT charge and mass in order to get the appropriate differential form of the thermodynamic first law recently \cite{Hennigar:2019ive,Bordo:2019slw, BallonBordo:2019vrn,Durka:2019ajz,Wu:2019pzr, Chen:2019uhp,BallonBordo:2020mcs, Awad:2020dhy,Abbasvandi:2021nyv,Frodden:2021ces,Wu:2022rmx,Godazgar:2022jxm, Awad:2022jgn, Wu:2022mlz,Wu:2022xpp,Wu:2023woq,Liu:2022wku,Liu:2023uqf}. Some specific choices of thermodynamics have been used in the holographic study successfully\cite{Jiang:2019yzs,Chen:2023eio,Perry:2022udk,Siahaan:2022jrl}. And recent results argue that the ill-defined thermodynamics can break weak cosmic supervision conjecture \cite{Yang:2023hll}, although a more careful examination on the subject is required \cite{Wu:2024ucf}. In this paper, we adopt the thermodynamics developed in \cite{Liu:2022wku} mainly for the following three reasons
\begin{itemize}

\item The mass, or the energy, of the black hole is always positive.

\item  The Euclidean action gives rise to the Gibbs free energy, $G=M- T S - \Phi_n Q_n$, analogous to the Kerr and Reissner-Nordstr\"om black holes.

\item There exists a smooth $n\rightarrow 0$ limit.
\end{itemize}
Arguably, the last criterium may be the weakest requirement.

In the framework of quantum field theory, Einstein theory coupled to conventional matter fields is UV divergent \cite{tHooft:1974toh} and non-renormalizable. A consistent quantum theory of gravity is still beyond reach. One way to study the higher energy physics at low energy scale is to consider the effective field theory approach. In the premise of preserving the covariance of the gravitational theory, the only choice of the low energy effective action of quantum gravity is adding all possible diffeomorphism invariant curvature terms \cite{Stelle:1976gc} which is given by
 \begin{equation}
     \mathcal{L}=R-2\Lambda_0+\frac{1}{M_{p}^2}\mathcal{L}_{4}+\frac{1}{M_{p}^4}\mathcal{L}_{6}+....
 \end{equation}
where $M_{p}$ is Planck mass, which is related to the UV behavior and $\mathcal{L}_{i}$ denotes the combination of all the $i$'th order derivative terms. If we treat the higher-order terms as part of classical action, they will generally introduce additional massive modes including ghost modes. These modes can give rise to new black holes with additional hair \cite{Lu:2015cqa}.  Indeed, some new Taub-NUT solutions were constructed numerically in a few recent papers \cite{Bueno:2018uoy,Brihaye:2018bgc,Ibadov:2021oqf,Butler:2023tyt,
Chen:2024hsh,Mukherjee:2021erg,Mukherjee:2020lld}.  In the perturbative effective theory approach, on the other hand, there will be no new modes and hence no new hair, but the original Ricci-flat Taub-NUT solution can be modified by the perturbation. For the Taub-NUT black hole that has $S^2$ symmetry even with the time bundle, the perturbation can be explicitly solved. However, for the complicated Kerr Taub-NUT, an exact solution, even at the perturbative level, is unlikely to be found. This provides a serious challenge to discuss the higher-derivative correction to the black hole thermodynamics. Recently, Reall and Santos developed a technique for evaluating the leading-order correction to black hole thermodynamics without perturbative solutions \cite{Reall:2019sah} based on the Euclidean action. The RS method has been very successful in the case of asymptotically flat and AdS black holes. (For the latter more complicated case, see, e.g.~\cite{Reall:2019sah,Xiao:2022auy,Ma:2023qqj}.) However, Taub-NUT or Kerr Taub-NUT metrics are neither asymptotically flat nor AdS. In this paper, we would like to test the RS method using the simpler Taub-NUT metric and then apply the method directly on the Kerr-Taub-NUT.

The paper is organized as follows: In section 2, we review Taub-NUT black hole and its thermodynamics obtained in \cite{Liu:2022wku}. In Section 3, we apply the RS method to obtain the corrected thermodynamics of Taub-NUT black hole. In Section 4, we construct the first-order perturbative solution of Taub-NUT and adopt the standard Euclidean action method to evaluate the corrected thermodynamics. Then we compare them with the results obtained by the RS method and prove that the RS method is applicable in Taub-NUT. In Section 5, we use the RS method to obtain the leading-order correction of the Kerr Taub-NUT black hole thermodynamics. Finally, we conclude the paper in Section 6. In the appendix, we consider perturbations to other proposals of the Taub-NUT thermodynamics.

\section{Thermodynamics of Taub-NUT black hole}

In this section, we briefly review the thermodynamics of the Taub-NUT black hole in Einstein gravity. As mentioned earlier, there are many proposals for the Taub-NUT black hole thermodynamics. In this paper, we adopt the proposal of \cite{Liu:2022wku} for the reasons explained in the introduction.

We begin with the Ricci-flat Taub-NUT black hole metric
 \begin{equation}\label{ansatz}
     ds^2=-h\,(dt+2n\cos\theta d\phi)^2+\frac{dr^2}{f}+(r^2+n^2)(d\theta^2+\sin^2\theta d\phi^2)\,,
 \end{equation}
 where
\begin{equation}\label{f0}
     f(r)=h(r)=f_0(r)\equiv\frac{r^2-2mr-n^2}{r^2+n^2}\,.
 \end{equation}
The metric has two parameters, $m$ and $n$. When $n=0$, the metric reduces to the Schwarzschild black hole of mass $m$, which has an event horizon for positive mass $m$. When $n$ is turned on, the event horizon continue to exist and its location $r_0$ is given by the largest positive root of $f$. The corresponding entropy and temperature can be assigned as
\begin{equation}
    T=\frac{\sqrt{f'(r_{0})h'(r_{0})}}{4\pi}=\frac{1}{4\pi r_{0}}\,,\qquad S=\frac{\rm Area}{4}=\pi(r_{0}^2+n^2)\,.
\end{equation}
However, a new feature emerges when $n$ is nonzero: an event horizon exists regardless the value and the sign of $m$, be it positive, negative or zero. Therefore, as was pointed out in \cite{Liu:2022wku}, it is no longer sensible to treat $m$, a quantity unbounded below, as the mass. Instead, based on the generalized Komar integration, the mass is shown to be
\be
M=m+\frac{n^2}{r_{0}}=\sqrt{m^2+n^2}\,,
\ee
which is always positive regardless the value of $m$. The NUT charge $Q_n$ and its thermodynamic conjugate $\Phi_n$ are given by
\be
Q_{n}=\frac{n}{r_{0}}\,,\qquad \Phi_n = \ft12 n\,.
\ee
In particular, the NUT potential is determined from the degenerate Killing vector at the north and south pole, {\it i.e.}
\begin{equation}
    \ell_{\pm}=\partial_{\phi}\mp 2n\partial_{t}=\partial_{\phi}\mp 4\Phi_{n}\partial_{t}\,.\label{killingv}
\end{equation}
This definition of $\Phi_n$ becomes more apparent in the Kerr Taub-NUT black hole, as was illustrated in \cite{Liu:2022wku}. The thermodynamic quantities satisfy the first law \eqref{firstlaw}. Furthermore, the Euclidean action $I$, expressed in terms of the above set of thermodynamic quantities, give precisely the Gibbs free energy, namely
\begin{equation}
     F_G=\frac{I}{\beta}=M-TS-\Phi_{n}Q_{n}\,,
\end{equation}
which is inline with other well established black holes, e.g.~Kerr and RN black holes, whose Euclidean actions all give rise to the Gibbs free energy.

\section{Thermodynamic corrections from the RS method}

In this section, we adopt the RS method \cite{Reall:2019sah} to obtain the leading-order higher-derivative correction to the Taub-NUT black hole. The simplest higher derivative extension to Einstein gravity is to consider quadratic invariants of Riemann tensor, including $R^2$, $R^{\mu\nu} R_{\mu\nu}$ and $R^{\mu\nu\rho\sigma} R_{\mu\nu\rho\sigma}$ terms. However, in four dimensions, the Gauss-Bonnet combination is a total derivative, and furthermore, there can be the field redefinition $g_{\mu\nu} \rightarrow g_{\mu\nu} + c_1 R_{\mu\nu} + c_2 R g_{\mu\nu}$ that does not alter the physics perturbatively. Therefore, the nontrivial perturbative extension requires at least the cubic invariants. Using the appropriate field redefinition \cite{Endlich:2017tqa,Cano:2019ore}, we have
\bea
&&{\cal L}_6 =  \sqrt{-g}\Big(\alpha_1 {\rm Riem}_3 + \alpha_2 \widetilde {\rm Riem}_3\Big)\,,\nn\\
{\rm Riem}_3 &=& R_{\mu\nu}{}^{\rho\sigma}R_{\rho\sigma}
{}^{\alpha\beta}R_{\alpha\beta}{}^{\mu\nu}\,,\qquad
\widetilde {\rm Riem}_3 = R^{\mu}{}_\nu{}^\rho{}_\sigma\,
R^{\nu}{}_\alpha {}^\sigma{}_\beta\,R^{\alpha}{}_\mu {}^\beta{}_\rho\,.
\eea
(Here, we do not consider the parity odd combinations that involve one four-index epsilon tensor.) In fact, since the Taub-NUT metric is Ricci flat, the terms involving two Ricci tensor and scalars will naturally drop out also. Note that the third-order Lovelock combination vanishes in four dimensions, it follows that for Ricci-flat metrics, we must have
\be
{\rm Riem}_3=2\widetilde {\rm Riem}_3\,.
\ee
We therefore only need to consider the total Lagrangian
\be
{\cal L}_{\rm tot} = \sqrt{-g} (R + \alpha\, {\rm Riem}_3)\,,\label{lag}
\ee
where $\alpha$ is the coupling constant of the cubic Riemann tensor invariant.
The on-shell Euclidean action is thus given by
\be
I_{E} = \fft{1}{16\pi} \int_{\rm bulk} {\cal L}_{\rm tot} + I_{\rm surf}\,.
\ee
The surface term associated with the leading-order Einstein-Hilbert action is the well-known Gibbon-Hawking-York term. The term associated with the cubic invariant can be complicated, but as we shall see presently, it gives no contributions owing to its fast falloff in the asymptotic region.

According to the RS method, the total Euclidean action, up to and including the linear $\alpha$ order can be evaluated on the background Ricci-flat metric $g_0$, namely
\be
I_{\rm E} = I_0(g_0) + I_1(g_0) + {\cal O}(\alpha^2)\,,\label{RS0}
\ee
where $I_0(g_0)$ is Euclidean action of the Einstein theory and $I_1(g_0)$ is given by
\be
I_1(g_0) = \fft{\alpha}{16\pi} \int_{\rm bulk} {\rm Riem}_3\Big|_{g\rightarrow g_0}\,.
\ee
It is worth noting that for the Taub-NUT metric, we have
\be
{\rm Riem}_3 = \fft{96m^3-3m n^2}{r^9} + {\cal O}(r^{-10})\,,
\ee
which is highly convergent asymptotically.

As we have pointed out in the previous section, in the thermodynamic interpretation of \cite{Liu:2022wku} we adopted, the Euclidean action is associated with the Gibbs free energy, where the two independent thermodynamic variables are $T$ and $\Phi_n$. Therefore, in the RS method \eqref{RS0}, these two variables remain their unperturbed values $(T_0, \Phi_{n0})$, namely
\begin{equation}
 T\rightarrow T_{0}=\frac{1}{4\pi \tilde{r}_{0}}+\mathcal{O}(\alpha^2)\,,
\qquad\Phi_{n}\rightarrow\Phi_{n0}=\frac{\tilde{n}}{2}+\mathcal{O}(\alpha^2)\,.
\end{equation}
For the later purpose of comparing the results of finding explicit perturbative solutions, we find it useful add tildes on parameters $r_0$ and $n$, which can be {\it a priori} perturbed by the cubic Riemann invariant. We find
 \begin{equation}
I_{0}(g_{0})=\frac{\tilde{r}_{0}^2-\tilde{n}^2}{4\tilde{r}_{0} T_0}\,,
\qquad I_{1}(g_{0})=\frac{5\tilde{n}^2-7\tilde{r}_{0}^2}{14\tilde{r}_{0}^3
    (\tilde{r}_{0}^2+\tilde{n}^2)T_0}\,.\label{RSres}
\end{equation}
Then, the corrected Gibbs free energy is
\begin{equation}
    G_{\rm RS}(T_{0},\Phi_{n0},\alpha)= T_{0}I_{E}(T_{0},\Phi_{n0},\alpha)=
\frac{\tilde{r}_{0}^2-\tilde{n}^2}{4\tilde{r}_{0}}+
\alpha\frac{(5\tilde{n}^2-7\tilde{r}_{0}^2)}{14\tilde{r}_{0}^3(\tilde{r}_{0}^2
+\tilde{n}^2)}.
\end{equation}
Note that we have omitted the ``${\cal O}(\alpha^2)$'' expression for simplicity. Entropy, NUT charge can be computed by the partial derivative of free energy and mass can be obtained by using a Legendre transformation
\begin{equation}
    S_{\rm RS}=-(\frac{\partial G_{\rm RS}}{\partial T_{0}})|_{\Phi_{n}}\,,\qquad Q_{n\rm RS}=-(\frac{\partial G_{\rm RS}}{\partial \Phi_{n0}})|_{T}\,,\qquad M_{\rm RS}=G_{\rm RS}+T_{0}S_{\rm RS}+\Phi_{n0}Q_{n\rm RS}\,.
\end{equation}
Specifically, we have
\bea
S_{\rm RS}&=&\pi (\tilde{r}_{0}^2+\tilde{n}^2)-\alpha\frac{6\pi(5\tilde{n}^4+6\tilde{n}^2\tilde{r}_{0}^2
-7\tilde{r}_{0}^4)}{7\tilde{r}_{0}^2(\tilde{r}_{0}^2+\tilde{n}^2)^2}\,,\nn\\
Q_{n\rm RS}&=& \frac{\tilde{n}}{\tilde{r}_{0}}-\alpha\frac{24 \tilde{n}}{7\tilde{r}_{0}(\tilde{r}_{0}^2+\tilde{n}^2)^2}\,,\qquad
M_{\rm RS} = \frac{\tilde{r}_{0}^2+\tilde{n}^2}{2\tilde{r}_{0}}-
\alpha\frac{5\tilde{n}^4+22\tilde{n}^2\tilde{r}_{0}^2
-7\tilde{r}_{0}^4}{7\tilde{r}_{0}^3(\tilde{r}_{0}^2+\tilde{n}^2)^2}\,.\label{mrs}
\eea

\section{Higher derivative correction to Taub-NUT metric}

As mentioned in Introduction,  Taub-NUT spacetime is not asymptotically Minkowskian but only locally flat. It is thus instructive to verify the correction of the cubic Riemann invariant from the more traditional approach. In this section we shall solve the equations of motion and construct the first-order perturbative solution to the Ricci-flat Taub-NUT and use the traditional method to compute the thermodynamic quantities.

\subsection{Perturbative solution}

The equation of motion (EOM) that follows from the theory \eqref{lag} can be conveniently written as
\begin{equation}
    P_{acde}R_{b}^{~cde}-\ft{1}{2}g_{ab}L-2\nabla^{c}\nabla^{d}P_{acdb}=0,
\end{equation}
where
\begin{equation}\label{pabcd}
         P_{abcd}=\frac{\partial L}{\partial R^{abcd}}=\ft12(g_{ac}g_{bd}-g_{ad}g_{bc})+
         \ft32\alpha(R_{a~c}^{~e~f}R_{bedf}-R_{a~d}^{~e~f}R_{becf})\,.
\end{equation}
We now move on to consider the ansatz (\ref{ansatz}) of Taub-NUT black hole. We wish to construct perturbative solutions to the first order in $\alpha$. We therefore consider metric functions in small $\alpha$ expansion that gives
\begin{equation}
f(r)=f_{0}(r)+\alpha f_{1}(r)+\mathcal{O}(\alpha^2)\,,\qquad
h(r)=f_{0}(r)+\alpha h_{1}(r)+\mathcal{O}(\alpha^2)\,.
\end{equation}
Here $f_{0}$ was given in \eqref{f0}. We substitute the expressions into the EOM and extract the linear differential equations of $ f_{1}$ and $ h_{1}$ at the linear $\alpha$ order. In order to simplify these equations, we first utilize the $rr$ component of EOM that yields
\begin{equation}
    \begin{split}
        f_{1}= &(-(r^2+n^2)^7(n^4+n^2(4m-5r)r-2mr^3)h_{1}+(n^2+2mr-r^2)(-24(-6m n^2r
        \\&(9n^8-60n^6r^2-42n^4r^4+132n^2r^6-23r^8)+2m^3(27n^8r-84n^6r^3-126n^4r^5
        \\&+108n^2r^7-5r^9)+n^2(n^{10}+27n^8r^2-126n^6r^4-210n^4r^6+189n^2 r^8-9r^{10})
        \\&+3m^2(5n^{10}-105n^8r^2+42n^6r^4+294n^4r^6-111n^2r^8+3r^{10}))+r(r^2+n^2)^8
        \\&h_{1}'))/(r(r^2+n^2)^7(-2mn^2+3n^2r+r^3))\,.
    \end{split}
\end{equation}
The $tt$ component of EOM then gives a second order differential equation of $h_{1}$:
\begin{equation}
    \begin{split}
        &h_1''+(48(8m^4n^2r^3(213n^8-924n^6r^2+846n^4r^4-188n^2r^6+5r^8)-48m^2n^2r^3
        \\&(106n^{10}-841n^8r^2+1176n^6r^4-190n^4r^6-122n^2r^8+15r^{10})+8n^4r^3(23 n^{10}
        \\&-674n^8r^2+1602n^6r^4-168n^4r^6-273n^2r^8+18r^{10})+m^3(15n^{14}+831n^{12}r^2
        \\&-15129n^{10}r^4+36399n^8r^6-18795n^6r^8-459n^4r^{10}+1013n^2r^{12}-35r^{14})+mn^2
        \\&(n^{14}-175n^{12}r^2+9225n^{10}r^4-41151n^8r^6+27867n^6r^8+6075n^4r^{10}-2661n^2r^{12}
        \\&+51r^{14}))+n^2(r^2+n^2)^8(4r^3+m(n^2-3r^2))h_{1}+r(n^2+r^2)^8(5n^2r^3+r^5-m(n^4
        \\&+5n^2r^2))h_{1}')/(r^2(n^2+r^2)^9(-2mn^2+3n^2r+r^3))=0\,.
    \end{split}
\end{equation}
The equation can be solved analytically.  We thus have
\begin{equation}
f_{1}=\Delta f_{1}+f_{bdy}\,,\qquad
h_{1}=\Delta h_{1}+h_{bdy}\,,
\end{equation}
where
\bea
\Delta f_{1}& =& \frac{8}{7r^2(r^2+n^2)^7} \Big[-3mn^2r(27n^8+2116n^6r^2-8598n^4r^4+6900 n^2r^6-973 r^8)\cr
&&+m^3(81n^8r+1036n^6r^3-10290n^4r^5+6972n^2r^7-343r^9)+3m^2(9 n^{10}\cr
&&+111n^8r^2-5298n^6r^4+10226n^4r^6-2871n^2r^8+63r^{10})-n^2(27n^{10}+669n^8r^2\cr
&&-5478n^6r^4+8390n^4r^6-3861n^2r^8+189r^{10})\Big].\\
\Delta h_{1}&=&\frac{8}{7r(r^2+n^2)^7}\Big[4n^4r(15n^6-237n^4r^2+265n^2r^4-27r^6)
-12m^2(33n^8r\cr
&&-219n^6r^3+247n^4r^5-45n^2r^7)-3mn^2(9n^8-404n^6r^2+1230n^4r^4\cr
&&-516n^2r^6+17r^8)+m^3(27n^8-476n^6r^2+1050n^4r^4-588n^2r^6+35r^8)\Big].
\eea
The functions $f_{bdy}$ and $h_{bdy}$ are terms associated with the integration constants $(c_1,c_2)$:
\bea
f_{bdy}&=&c_{1}\frac{2n^4+(-3+4m)r^3+3n^2r(2m+(-2 + r)r)}{(-1+(-3+2m)n^2)r^2(n^2+r^2)}\cr
&&+c_{2}\frac{-n^4+(1-2m)r^3+
n^2r(-3(-1+r)r+m(-3+r^2))}{(-1+(-3+2m)n^2)r^2(n^2+r^2)}\,,\\
h_{bdy}&=&-c_{1}\frac{2mn^2+r(3n^2(-2+r)+r(-3+2r))}{(r+(3-2m)n^2r)(n^2+r^2)}\cr
&&+c_{2}\frac{(-1+r)(-mn^2(1+r)+r(3n^2+r))}{(r+(3-2m)n^2r)(n^2+r^2)}\,.
\eea
The integration constants can be fixed in the following consideration. In the large $r$ region, $f_{1}$ and $h_{1}$ should vanish. We find
\begin{equation}\label{lim}
\lim_{r \to\infty}f_{1}=0\,,\qquad\lim_{r \to\infty}h_{1}=\frac{2c_{1}-c_{2}}{-1(-3+2m)n^2}\,,\qquad
\rightarrow\qquad c_2=2c_1\,.
\end{equation}
Then, we have
\begin{equation}
f_{1}=\frac{c_{1}}{r}+\mathcal{O}(\frac{1}{r^2}).
\end{equation}
The perturbative metric function $f_{1}$ should have a $[L]^{-4}$ dimension. It indicates that if $c_{1}\neq 0$, it would have the following form
\begin{equation}
  c_{1}=\frac{1}{\sum_{i=0}^{3}\beta_{i} m^i n^{3-i}}\,.
\end{equation}
where $\beta_{i}$ are some purely numerical constants. Obviously, the metric is divergent when $m=0$ and $n=0$. Therefore, $c_{1}$ must vanish, and our perturbative solution should be $f_{1}=\Delta f_{1}$ and $h_{1}=\Delta h_{1}$.

\subsection{Thermodynamics}

In our perturbative approach, the black hole horizon is no longer located at $r_0$ but corrected by higher derivative terms. It can be determined by $f(r_{h})=0=h(r_h)$. The horizon is thus shifted by (at the linear $\alpha$ order)
\begin{equation}
    r_{h}=r_{0}+\alpha\frac{27n^2-35r_{0}^2}{7n^2 r_{0}^3+7r_{0}^5}\,.
\end{equation}
The Hawking temperature can be evaluated by the standard method
\begin{equation}\label{T}
    T=\frac{\sqrt{f'(r_{h})h'(r_{h})}}{4\pi}=\frac{1}{4\pi r_{0}}+\alpha\frac{7r_{0}^2-5n^2}{14\pi r_{0}^3(r_{0}^2+n^2)^2}\,.
\end{equation}
In higher derivative gravity, the entropy is not simply given by one quarter of the area of the event horizon. Instead, we need to use the formula obtained by Wald's prescription \cite{Wald:1993nt,Iyer:1994ys}, namely
\begin{equation}
     S=\frac{1}{8\pi}\int d^2x\sqrt{h} \epsilon^{ab}\epsilon^{cd}P_{abcd}\,,
\end{equation}
where $h$ is determinant of the induced metric on the horizon, $\epsilon_{ab}$ is the binormal to the horizon with $\epsilon_{ab}\epsilon^{ab}=-2$ and $P_{abcd}$ is defined in (\ref{pabcd}). Specifically, we find the Wald entropy of our linearly perturbed solution is given by
\begin{equation}
    S=\pi(r_{0}^2+n^2)+\alpha\frac{2\pi(7r_{0}^2-15n^2)}{7r_{0}^2(r_{0}^2+n^2)}\,.
\end{equation}
The degenerate Killing vectors \eqref{killingv} at the north and south pole remain intact under the perturbation; therefore, the NUT potential remains the same
\begin{equation}
    \Phi_{n}=\frac{n}{2}=\Phi_{n0}\,.
\end{equation}
The remainder of the thermodynamic variables can be obtained from the free energy associated with the Euclidean action, namely
\begin{equation}
   G=\frac{I}{\beta}\,,\qquad Q_{n}=-(\frac{\partial G}{\partial \Phi_{n}})\Big|_{T}\,,\qquad M=F+TS+\Phi_{n}Q_{n}\,.
\end{equation}
Here we adopt the brutal-force method and simply substitute the perturbed solution into the Euclidean action, evaluate it up to and including the $\alpha$ order. Note that the surface term associated with the cubic invariant term can be ignored, owing to its fast falloff on the boundary. We find
\begin{equation}\label{F}
G=\frac{m}{2}+\alpha\frac{5n^2-7r_{0}^2}{7r_{0}^3(r_{0}^2+n^2)},
\end{equation}
and
\begin{equation}
 M=\frac{r_{0}^2+n^2}{2r_{0}}-\alpha\frac{10n^2}{7r_{0}^3(r_{0}^2+n^2)}\,,\qquad    Q_{n}=\frac{n}{r_{0}}-\alpha\frac{10n}{7r_{0}^3(r_{0}^2+n^2)}\,.
\end{equation}
It is straightforward to verify that they do obey the first law \eqref{firstlaw} up to and including the $\alpha$ order. It is worth noting that there exists a relation between $\Delta Q_{n}$ and $\Delta M$:
 \begin{equation}
     \Delta M=n\Delta Q_{n}=2\Phi_{n}\Delta Q_{n}.
 \end{equation}
It indicates that the mass can be conveniently written as an $\alpha$-independent expression
 \begin{equation}\label{mq}
     M=m+2\Phi_{n}Q_{n}\,,
 \end{equation}
even under the $\alpha$-order correction.

\subsection{Verification of The RS Method in Taub-NUT}

We have obtained the $\alpha$-order correction to the Taub-NUT black hole thermodynamics using two different methods. We now show that they are equivalent.
We note that in the RS approach, the thermodynamic quantities are expressed in terms of $(\tilde r_0, \tilde n)$, such that the thermodynamic variables $(T, \Phi)$ are fixed to $T_{0}$ and $\Phi_{0}$. We therefore need to perform redefinition on $(r_0, n)$ in the second approach
\begin{equation}
     r_{0}\rightarrow \tilde{r}_{0}+\delta \tilde{r}_{0}\,,\qquad \delta\tilde{r}_{0}=-\alpha\frac{2(5\tilde{n}^2-7\tilde{r}_{0}^2)}
     {7\tilde{r}_{0}(\tilde{r}_{0}^2+\tilde{n}^2)^2}\,,\qquad n\rightarrow \tilde{n}\,,
 \end{equation}
so that the temperature and NUT potential are both unperturbed, namely
 \begin{equation}
     T\rightarrow T_{0}=\frac{1}{4\pi \tilde{r}_{0}}+\mathcal{O}(\alpha^2)\,,\qquad \Phi_{n}\rightarrow \Phi_{n0}+\mathcal{O}(\alpha^2)\,.
 \end{equation}
It is straightforward to verify that the second approach would reproduce the RS thermodynamic quantities precisely. We therefore establish the RS method is indeed applicable in Taub-NUT geometries.

It is also of interest to consider thermodynamic system where the entropy is the potential, in which case, the mass and NUT charge are the thermodynamic variables that are unperturbed under the $\alpha$ correction. Starting from the second approach, we need perform a further field redefinition
\be
r_{0}=\hat{r}_{0}+\alpha \delta r_{0}\,,\qquad n\rightarrow \hat{n}+\alpha\delta n\,;\qquad
\delta r_{0} = 0\,,\qquad
\delta n=\frac{10\hat{n}}{7\hat{r}_{0}^2(\hat{r}_{0}^2+\hat{n}^2)}\,,
\ee
under which the mass and NUT charge are both fixed
\begin{equation}
    M=\frac{\hat{r}_{0}^2+\hat{n}^2}{2\hat{r}_{0}}+\mathcal{O}(\alpha^2)\,,\qquad
    Q_{\hat{n}}=\frac{n}{\hat{r}_{0}}+\mathcal{O}(\alpha^2)\,.
\end{equation}
The linear $\alpha$ correction of the remaining thermodynamic quantities are
\bea
T &=&\frac{1}{4\pi \hat{r}_{0}}+\frac{\alpha(7\hat{r}_{0}^2-5\hat{n}^2)}{14\hat{r}_{0}^3
(\hat{r}_{0}^2+\hat{n}^2)^2}\,,\quad
S=\pi(\hat{r}_{0}^2+\hat{n}^2)+\frac{2\pi\alpha(7\hat{r}_{0}^2
-5\hat{n}^2)}{7\hat{r}_{0}^2(\hat{r}_{0}^2+\hat{n}^2)}\,,\cr
\Phi_{n} &=& \frac{\hat{n}}{2}+\frac{5\alpha\hat{n}}{7\hat{r}_{0}^2
(\hat{r}_{0}^2+\hat{n}^2)}\,.
\eea
We thus reestablish the identity \cite{Reall:2019sah}
\begin{equation}
    \Delta S\Big|_{M,Q_{n}}=\frac{2\pi\alpha(7\hat{r}_{0}^2
    -5\hat{n}^2)}{7\hat{r}_{0}^2(\hat{r}_{0}^2+\hat{n}^2)}=-\Delta I\Big|_{T,\Phi_{n}}\,.
\end{equation}
If we instead start from the RS method, we can get the same result by performing the field redefinition
\bea
\tilde{r}_{0}=\hat{r}_{0}+\alpha \delta \tilde{r}_{0}\,,\quad\tilde{n}\rightarrow \hat{n}+\alpha\delta \tilde{n}\,;\qquad
\delta\tilde{r}_{0}=\frac{2(5\hat{n}^2-7\hat{r}_{0}^2)}{7\hat{r}_{0}^2(\hat{r}_{0}^2
+\hat{n}^2)^2}\,,
\quad \delta \tilde{n}=\frac{10\hat{n}}{7\hat{r}_{0}^2(\hat{r}_{0}^2+\hat{n}^2)}\,.
\eea

\subsection{Comparing the two approaches}

In both approaches described in this section, we obtain the correction to the Gibbs free energy, from which we read off the thermodynamic quantities. However, the Gibbs free energy is expressed in terms of the parameters associated with the horizon radius and the NUT parameter $n$. By itself, we cannot read off the thermodynamic quantities directly. In the RS approach, the thermodynamic variables are {\it assumed} to be fixed, without corrections. We can then read off the remaining thermodynamic quantities associated with the differential relation from $dG = - S dT - Q_n d\Phi_n$. In the approach of constructing explicit perturbative solution, the issue becomes more subtle. Since the explicit solution is given, we can determine the temperature and entropy of the perturbed solution by the standard methods, which leads to $Q_n d\Phi_n = - S dT-dG$. However, we cannot determine $(\Phi_n, Q_n)$ simply from $Q_n d\Phi_n$. We need an independent formula to determine either $\Phi_n$ or $Q_n$. The key of the thermodynamic approach of \cite{Liu:2022wku} is that the NUT potential is determined by \eqref{killingv}. Since the degenerate Killing vectors remain the same for our perturbative solution, it follows that $\Phi_n$ is unperturbed in our second approach. We can then read off the perturbed NUT charge $Q_n$ from the perturbed Gibbs free energy.

Thus we see that in the both RS and standard method, the Gibbs free energy can be determined, but one needs to know independently the $(T, \Phi_n)$ in order to read off the complete set of thermodynamic variables. In the RS case, by procedure, we have $(T, \Phi_n)=(T_0, \Phi_{n0})$. For the standard method of constructing perturbative solution, $T$ is no longer $T_0$, but it can be obtained by the standard procedure, and $\Phi_n$ happens to be $\Phi_{n0}$. Thus it is a highly nontrivial exercise that the RS and perturbative solution methods match precisely and it provides an important validation to the thermodynamic proposal of \cite{Liu:2022wku}.

In the appendix, we consider some other thermodynamic proposals. One can always apply the RS formalism, but its verification using the perturbative solution becomes less convincing, since there is no clear independent calculation of the thermodynamic variables from the perturbative solution.

\section{First-order correction to Kerr-Taub-NUT thermodynamics}

Having established that the RS method is applicable to the Taub-NUT black hole, we now apply it to the Kerr Taub-NUT black hole, for which the analytic perturbative solution is unlikely to be found. The Ricci-flat rotating Taub-NUT metric is given by
\begin{equation}
\begin{split}
    &ds^2=(r^2+v^2)(\frac{dr^2}{\Delta}+\frac{du^2}{1-u^2})
    +\frac{1}{r^2+v^2}((1-u^2)e_{1}^2-\Delta e_{2}^2)\,,
    \\&e_{1}=adt-(r^2+a^2+n^2)d\phi\,,\qquad e_{2}=dt+(2nu-a(1-u^2))d\phi\,,
    \\&\Delta=r^2-2mr+a^2-n^2\,,\qquad v=n+au\,,\qquad u=\cos\theta.
\end{split}
\end{equation}
The metric consists of three integration constants $(m,a,n)$. For $n=0$, it reduces to the standard Kerr metric. The metric contains a Cauchy horizon $r_-$ and an event horizon $r_+\ge r_-$, which are defined by $\Delta(r_{\pm})=0$. The thermodynamic quantities based on \cite{Liu:2022wku} are:
\begin{equation}
    \begin{split}
        &T=\frac{r_{+}^2+n^2-a^2}{4\pi r_{+}(r_{+}^2+n^2+a^2)}\,,\qquad\Omega=\frac{a}{r_{+}^2+n^2+a^2}\,,
        \qquad \Phi_{n}=\frac{n}{2}\,,
        \\&S=\pi(r_{+}^2+n^2+a^2)\,,\qquad Q_{n}=\frac{n}{r_{+}}~,~~M=m+\frac{n^2}{r_{+}}\,,\qquad J=Ma\,.
    \end{split}
\end{equation}
Again, the free energy associated with the Euclidean action is of the Gibbs type:
\be
G(T,\Phi_n, \Omega)=M - T S - \Phi_n Q_n - \Omega J\,.
\ee
We now apply the RS method. We choose the redefinition of integration constant $r_{+}\rightarrow \tilde{r_0}$, $a\rightarrow \tilde{a}$, $n\rightarrow \tilde{n}$. The NUT parameter is also not redefined as we discussed in the last section. We have
\bea
    I_{0}(g_{0}) &=& \frac{\tilde{r}_{0}^2-\tilde{n}^2+
    \tilde{a}^2}{4\tilde{r}_{0}}\beta\,,\nn\\
    I_{1}(g_{0})&=&\frac{\beta}{14\tilde{r_0}^3(\tilde{a}^2-
    2\tilde{a}\tilde{n}+\tilde{n}^2+\tilde{r_0}^2)^3(\tilde{a}^2+2\tilde{a}\tilde{n}+\tilde{n}^2
    +\tilde{r_0}^2)^3}\Big[\tilde{a}^{12}+6\tilde{a}^{10}(\tilde{n}^2-3\tilde{r}_{0}^2)\cr
&&+(5\tilde{n}^2-7\tilde{r}_{0}^2)(\tilde{n}^2+\tilde{r}_{0}^2)^5- 2\tilde{a}^2(\tilde{n}^2+\tilde{r}_{0}^2)^3(5\tilde{n}^4+78\tilde{n}^2\tilde{r}_{0}^2-7\tilde{r}_{0}^4)\cr
&&-\tilde{a}^8(29\tilde{n}^4+198\tilde{n}^2\tilde{r}_{0}^2+25\tilde{r}_{0}^4)
+12\tilde{a}^6(3\tilde{n}^6+25\tilde{n}^4\tilde{r}_{0}^2+25\tilde{n}^2\tilde{r}_{0}^4
+3\tilde{r}_{0}^6)\cr
&&+\tilde{a}^4(-9\tilde{n}^8+84\tilde{n}^6\tilde{r}_{0}^2+450\tilde{n}^4\tilde{r_0}^4
+420\tilde{n}^2\tilde{r_0}^6+63\tilde{r_0}^8)\Big].
\eea
The Gibbs-type free energy is then simply $G=(I_{0}(g_{0})+ I_{1}(g_{0}))T_{0}$. we can obtain the perturbed entropy, angular momentum, NUT charge and mass:
\begin{equation}
\begin{split}
    &S=-(\frac{\partial F}{\partial T_{0}})|_{\Phi_{n},\Omega}~,~~J=-(\frac{\partial F}{\partial \Omega_{0}})|_{\Phi_{n},T}~,~~Q_{n}=-(\frac{\partial F}{\partial \Phi_{n0}})|_{T,\Omega},
    \\&M=F+T_{0}S+\Phi_{n0}Q_{n}+\Omega_{0}J.
\end{split}
\end{equation}
Specially, the corrections are
\begin{equation}
    \begin{split}
        \Delta S=&\alpha(2\pi(\tilde{a}^2+\tilde{n}^2+\tilde{r_0}^2)(3\tilde{a}^{14}+3\tilde{a}^{12}(5\tilde{n}^2-3\tilde{r_0}^2)-3(\tilde{n}^2+\tilde{r_0}^2)^5(5\tilde{n}^4+6\tilde{n}^2\tilde{r_0}^2-
        7\tilde{r_0}^4)\\&-\tilde{a}^{10}(105\tilde{n}^4+150\tilde{n}^2\tilde{r_0}^2+101\tilde{r_0}^4)+\tilde{a}^8(195\tilde{n}^6+469\tilde{n}^4\tilde{r_0}^2-919\tilde{n}^2\tilde{r_0}^4-233\tilde{r_0}^6)\\&+\tilde{a}^2(\tilde{n}^2+\tilde{r_0}^2)^3(45\tilde{n}^6+179\tilde{n}^4\tilde{r_0}^2+503\tilde{n}^2\tilde{r_0}^4+49\tilde{r_0}^6)-3\tilde{a}^6(45\tilde{n}^8
        +108\tilde{n}^6\tilde{r_0}^2
        \\&+150\tilde{n}^4\tilde{r_0}^4+284\tilde{n}^2\tilde{r_0}^6+69\tilde{r_0}^8) +\tilde{a}^4(-3\tilde{n}^{10}-207\tilde{n}^8\tilde{r_0}^2+514\tilde{n}^6\tilde{r_0}^4+1218\tilde{n}^4\tilde{r_0}^6\\&+465\tilde{n}^2\tilde{r_0}^8-35\tilde{r_0}^{10})))/(7\tilde{r_0}^2(\tilde{a}^2-2\tilde{a}\tilde{n}+\tilde{n}^2+\tilde{r_0}^2)^4(\tilde{a}^2+2\tilde{a}\tilde{n}+\tilde{n}^2+\tilde{r_0}^2)^4)\,,
    \end{split}
\end{equation}
\begin{equation}
    \begin{split}
        \Delta J=&-\alpha((2\tilde{a}(\tilde{a}^2+\tilde{n}^2+\tilde{r_0}^2)(9\tilde{a}^{12}(\tilde{n}^2-\tilde{r_0}^2)+2\tilde{a}^{10}(5\tilde{n}^4-126\tilde{n}^2\tilde{r_0}^2-11\tilde{r_0}^4)\\&+ (\tilde{n}^2+\tilde{r_0}^2)^5(25\tilde{n}^4-96\tilde{n}^2\tilde{r_0}^2+7\tilde{r_0}^4)+4\tilde{a}^6(\tilde{n}^2+\tilde{r_0}^2)^2(35\tilde{n}^4+198\tilde{n}^2\tilde{r_0}^2
        \\&+19\tilde{r_0}^4)- \tilde{a}^4(\tilde{n}^2+\tilde{r_0}^2)^2(25\tilde{n}^6-407\tilde{n}^4\tilde{r_0}^2-809\tilde{n}^2\tilde{r_0}^4-89\tilde{r_0}^6)-2\tilde{a}^2(\tilde{n}^2
        \\&+\tilde{r_0}^2)^3(27\tilde{n}^6+281\tilde{n}^4\tilde{r_0}^2- 23\tilde{n}^2\tilde{r_0}^4-21\tilde{r_0}^6)-\tilde{a}^8(105\tilde{n}^6+473\tilde{n}^4\tilde{r_0}^2+71\tilde{n}^2\tilde{r_0}^4
        \\&-9\tilde{r_0}^6)))/(7\tilde{r_0}^3(\tilde{a}^2-2\tilde{a}\tilde{n}+\tilde{n}^2+ \tilde{r_0}^2)^4(\tilde{a}^2+2\tilde{a}\tilde{n}+\tilde{n}^2+\tilde{r_0}^2)^4))\,,
    \end{split}
\end{equation}
\begin{equation}
    \begin{split}
        \Delta Q_n=&-(4\alpha \tilde{n}(9\tilde{a}^{14}+10\tilde{a}^{12}(\tilde{n}^2-18\tilde{r_0}^2)+6\tilde{r_0}^2(\tilde{n}^2+\tilde{r_0}^2)^6+5\tilde{a}^2(\tilde{n}^2+\tilde{r_0}^2)^4(5\tilde{n}^4
        \\&-2\tilde{n}^2\tilde{r_0}^2-23\tilde{r_0}^4)-\tilde{a}^{10}(105\tilde{n}^4+766\tilde{n}^2\tilde{r_0}^2+181\tilde{r_0}^4)-2\tilde{a}^4(\tilde{n}^2+\tilde{r_0}^2)^2(27\tilde{n}^6
        \\&+406\tilde{n}^4\tilde{r_0}^2+411\tilde{n}^2\tilde{r_0}^4-16\tilde{r_0}^6)+2\tilde{a}^8 (70\tilde{n}^6+675\tilde{n}^4\tilde{r_0}^2+900\tilde{n}^2\tilde{r_0}^4+247\tilde{r_0}^6)
        \\&+\tilde{a}^6(-25\tilde{n}^8+420\tilde{n}^6\tilde{r_0}^2+2130\tilde{n}^4\tilde{r_0}^4+2324\tilde{n}^2\tilde{r_0}^6+639\tilde{r_0}^8)))/(7\tilde{r_0}^3(\tilde{a}^2-2\tilde{a}\tilde{n}
        \\&+\tilde{n}^2+\tilde{r_0}^2)^4(\tilde{a}^2+2\tilde{a}\tilde{n}+\tilde{n}^2+\tilde{r_0}^2)^4)\,,
    \end{split}
\end{equation}
\begin{equation}
    \begin{split}
        \Delta M=&-(\alpha(\tilde{a}^{16}+8\tilde{a}^{14}(5\tilde{n}^2-2\tilde{r_0}^2)+16\tilde{a}^2\tilde{n}^2(\tilde{n}^2+\tilde{r_0}^2)^4(5\tilde{n}^4-12\tilde{n}^2\tilde{r_0}^2-37\tilde{r_0}^4)
        \\&+(\tilde{n}^2+\tilde{r_0}^2)^6(5\tilde{n}^4+22\tilde{n}^2\tilde{r_0}^2-7\tilde{r_0}^4)-4\tilde{a}^{12}(217\tilde{n}^2\tilde{r_0}^2+15\tilde{r_0}^4)-32\tilde{a}^{10}(10\tilde{n}^6
        \\&+76\tilde{n}^4\tilde{r_0}^2+25\tilde{n}^2\tilde{r_0}^4+\tilde{r_0}^6)-4\tilde{a}^4(\tilde{n}^2+\tilde{r_0}^2)^2(50\tilde{n}^8+711\tilde{n}^6\tilde{r_0}^2+723\tilde{n}^4\tilde{r_0}^4
        \\&-55\tilde{n}^2\tilde{r_0}^6-21\tilde{r_0}^8)+2\tilde{a}^8(225\tilde{n}^8+2350\tilde{n}^6\tilde{r_0}^2+3600\tilde{n}^4\tilde{r_0}^4+1338\tilde{n}^2\tilde{r_0}^6+55\tilde{r_0}^8)
        \\&+8\tilde{a}^6(-7\tilde{n}^{10}+210\tilde{n}^8\tilde{r_0}^2+1030\tilde{n}^6\tilde{r_0}^4+1232\tilde{n}^4\tilde{r_0}^6+441\tilde{n}^2\tilde{r_0}^8+22\tilde{r_0}^{10})))/(7\tilde{r_0}^3
        \\&(\tilde{a}^2-2\tilde{a}\tilde{n}+\tilde{n}^2+\tilde{r_0}^2)4(\tilde{a}^2+2\tilde{a}\tilde{n}+\tilde{n}^2+\tilde{r_0}^2)^4)\,.
    \end{split}
\end{equation}
Since in Gibbs-type free energy, the temperature is an independent variable. We can thus take (smooth) extremal limit by simply setting $T_0=0$, which implies that $\tilde{r}_{0}^2=\tilde{a}^2-\tilde{n}^2$. The results become significantly simpler, given by
\bea
S&=&2\pi\tilde{a}^2-\frac{4\pi\alpha(2\tilde{a}^2+3\tilde{n}^2)}{
7(\tilde{a}^2-\tilde{n}^2)^2}\,,\qquad
J=\frac{\tilde{a}^3}{\sqrt{\tilde{a}^2-\tilde{n}^2}}-\frac{\alpha \tilde{a}^3 (3 \tilde{a}^2 + 17 \tilde{n}^2)}{7 (\tilde{a}^2-\tilde{n}^2)^3 \sqrt{\tilde{a}^2 - \tilde{n}^2}}\,,\cr
Q_{n} &=& \frac{\tilde{n}}{\sqrt{\tilde{a}^2-\tilde{n}^2}}+\frac{\alpha \tilde{n} (11 \tilde{a}^2 + 9 \tilde{n}^2)}{7 (\tilde{n}^2-\tilde{a}^2)^3\sqrt{\tilde{a}^2 - \tilde{n}^2}}\,,\quad
M=\frac{\tilde{a}^2}{\sqrt{\tilde{a}^2-\tilde{n}^2}}+\frac{\alpha{(-\tilde{a}^4 - 13 \tilde{a}^2 \tilde{n}^2 - 6 \tilde{n}^4)}}{7 (\tilde{a}^2-\tilde{n}^2)^3\sqrt{\tilde{a}^2 - \tilde{n}^2}}.
\eea

Finally, we consider corrections with the charges, namely $(M,J,Q_{n})$, fixed.
This can be achieved by field redefinitions
\begin{equation}
    \tilde{r}_{0}\rightarrow \hat{r}_{0}+\alpha\delta\tilde{r}_{0}\,,\qquad \tilde{n}\rightarrow \hat{n}+\alpha\delta\tilde{n}\,,\qquad \tilde{a}\rightarrow \hat{a}+\alpha\delta\tilde{a}\,.
\end{equation}
where
\begin{equation}
    \begin{split}
        \delta\tilde{r}_{0}=&-2 (-3 \hat{a}^{18} - (5 \hat{n}^2 - 7 \hat{r}_{0}^2) (\hat{n}^2 + \hat{r}_{0}^2)^8 +
   \hat{a}^{16} (-49 \hat{n}^2 + 11 \hat{r}_{0}^2) +
   7 \hat{a}^2 (\hat{n}^2 + \hat{r}_{0}^2)^6 (15 \hat{n}^4
   \\&- 42 \hat{n}^2 \hat{r}_{0}^2 + 7 \hat{r}_{0}^4) +
   8 \hat{a}^{14} (14 \hat{n}^4 + 111 \hat{n}^2 \hat{r}_{0}^2 + 9 \hat{r}_{0}^4) -
   16 \hat{a}^4 (\hat{n}^2 + \hat{r}_{0}^2)^4 (17 \hat{n}^6 + 149 \hat{n}^4 \hat{r}_{0}^2
   \\&- 71 \hat{n}^2 \hat{r}_{0}^4 -
      7 \hat{r}_{0}^6) +
   8 \hat{a}^{12} (25 \hat{n}^6 + 195 \hat{n}^4 \hat{r}_{0}^2 + 63 \hat{n}^2 \hat{r}_{0}^4 + 13 \hat{r}_{0}^6) -
   2 \hat{a}^{10} (375 \hat{n}^8 + 3728 \hat{n}^6 \hat{r}_{0}^2
   \\&+ 4966 \hat{n}^4 \hat{r}_{0}^4 +
      1208 \hat{n}^2 \hat{r}_{0}^6 - 21 \hat{r}_{0}^8) +
   24 \hat{a}^6 (\hat{n}^2 + \hat{r}_{0}^2)^2 (\hat{n}^8 + 183 \hat{n}^6 \hat{r}_{0}^2 + 351 \hat{n}^4 \hat{r}_{0}^4
   \\&+109 \hat{n}^2 \hat{r}_{0}^6+ 4 \hat{r}_{0}^8) +
   2 \hat{a}^8 (319 \hat{n}^{10} + 1863 \hat{n}^8 \hat{r}_{0}^2 + 1190 \hat{n}^6 \hat{r}_{0}^4 -
      770 \hat{n}^4 \hat{r}_{0}^6 - 405 \hat{n}^2 \hat{r}_{0}^8
      \\&+ 11 \hat{r}_{0}^{10}))/(7 \hat{r}_{0} (\hat{a}^2 - 2 \hat{a} \hat{n} + \hat{n}^2 + \hat{r}_{0}^2)^4 (\hat{a}^2 + 2 \hat{a} \hat{n} + \hat{n}^2 +
     \hat{r}_{0}^2)^4 (-\hat{a}^4 + (\hat{n}^2 + \hat{r}_{0}^2)^2))\,,
    \end{split}
\end{equation}
\begin{equation}
    \begin{split}
        \delta \tilde{n}=&2 \hat{n} (-15 \hat{a}^{18} + 5 (\hat{n}^2 + \hat{r}_{0}^2)^9 + \hat{a}^{16} (29 \hat{n}^2 + 349 \hat{r}_{0}^2) +4 \hat{a}^{14} (29 \hat{n}^4 + 170 \hat{n}^2 \hat{r}_{0}^2 + 77 \hat{r}_{0}^4) -
        \\&\hat{a}^2 (\hat{n}^2 + \hat{r}_{0}^2)^6 (55 \hat{n}^4 - 274 \hat{n}^2 \hat{r}_{0}^2 + 279 \hat{r}_{0}^4) +
   4 \hat{a}^4 (\hat{n}^2 + \hat{r}_{0}^2)^4 (41 \hat{n}^6 + 187 \hat{n}^4 \hat{r}_{0}^2 - 701 \hat{n}^2 \hat{r}_{0}^4
   \\&- 15 \hat{r}_{0}^6) -4 \hat{a}^{12}(115 \hat{n}^6 + 1145 \hat{n}^4 \hat{r}_{0}^2 + 1201 \hat{n}^2 \hat{r}_{0}^4 + 363 \hat{r}_{0}^6) +
   2 \hat{a}^{10} (295 \hat{n}^8 + 2332 \hat{n}^6 \hat{r}_{0}^2+
   \\&1018 \hat{n}^4 \hat{r}_{0}^4 -
      2244 \hat{n}^2 \hat{r}_{0}^6 - 841 \hat{r}_{0}^8) -
   4 \hat{a}^6 (\hat{n}^2 + \hat{r}_{0}^2)^2 (31 \hat{n}^8 + 908 \hat{n}^6 \hat{r}_{0}^2 + 986 \hat{n}^4 \hat{r}_{0}^4
   \\&- 628 \hat{n}^2 \hat{r}_{0}^6- 353 \hat{r}_{0}^8) +2\hat{a}^8 (-125 \hat{n}^{10} + 687 \hat{n}^8 \hat{r}_{0}^2 + 5950 \hat{n}^6 \hat{r}_{0}^4 +
      8638 \hat{n}^4 \hat{r}_{0}^6 + 3951 \hat{n}^2 \hat{r}_{0}^8
      \\&+ 451 \hat{r}_{0}^{10}))/(7 \hat{r}_{0}^2 (\hat{a}^2 - 2 \hat{a} \hat{n} + \hat{n}^2 + \hat{r}_{0}^2)^4 (\hat{a}^2 + 2 \hat{a} \hat{n} + \hat{n}^2 +
    \hat{r}_{0}^2)^4 (-\hat{a}^4 + (\hat{n}^2 + \hat{r}_{0}^2)^2))\,,
    \end{split}
\end{equation}
\begin{equation}
    \begin{split}
        \delta \tilde{a}=&2 (-\hat{a}^{17}-2\hat{a}^{15}(11\hat{n}^2 + \hat{r}_{0}^2) +\hat{a}^{13} (38 \hat{n}^4 + 364 \hat{n}^2 \hat{r}_{0}^2 - 2 \hat{r}_{0}^4) + \hat{a} (\hat{n}^2 + \hat{r}_{0}^2)^6 (45 \hat{n}^4
        \\&- 214 \hat{n}^2 \hat{r}_{0}^2 + 21 \hat{r}_{0}^4) -
   2 \hat{a}^3 (\hat{n}^2 + \hat{r}_{0}^2)^4 (69 \hat{n}^6 + 537 \hat{n}^4 \hat{r}_{0}^2 - 253 \hat{n}^2 \hat{r}_{0}^4 -
      49 \hat{r}_{0}^6) +
   2 \hat{a}^{11} (65 \hat{n}^6
   \\&+ 501 \hat{n}^4 \hat{r}_{0}^2 + 55 \hat{n}^2 \hat{r}_{0}^4 + 3 \hat{r}_{0}^6) -4 \hat{a}^9 (95 \hat{n}^8 + 928 \hat{n}^6 \hat{r}_{0}^2 + 1172 \hat{n}^4 \hat{r}_{0}^4 + 228 \hat{n}^2 \hat{r}_{0}^6 -
      15 \hat{r}_{0}^8)
      \\&+2\hat{a}^5 (\hat{n}^2 + \hat{r}_{0}^2)^2 (21 \hat{n}^8 + 1188 \hat{n}^6 \hat{r}_{0}^2 +
      2146 \hat{n}^4 \hat{r}_{0}^4 + 876 \hat{n}^2 \hat{r}_{0}^6 + 89 \hat{r}_{0}^8) +2\hat{a}^7 (143 \hat{n}^{10}
      \\&+729 \hat{n}^8 \hat{r}_{0}^2 + 350 \hat{n}^6 \hat{r}_{0}^4 - 70 \hat{n}^4 \hat{r}_{0}^6 +
      243 \hat{n}^2 \hat{r}_{0}^8 + 77 \hat{r}_{0}^{10}))/(7 \hat{r}_{0}^2 (\hat{a}^2 + \hat{n}^2 + \hat{r}_{0}^2) (\hat{a}^2 - 2 \hat{a} \hat{n}
      \\&+ \hat{n}^2 +\hat{r}_{0}^2)^4 (\hat{a}^2 + 2 \hat{a} \hat{n}+ \hat{n}^2 + \hat{r}_{0}^2)^4)\,.
    \end{split}
\end{equation}
Under these redefinitions, the mass, NUT charge and angular momentum are fixed at the linear $\alpha$ order,
\begin{equation}
    M=M_{0}+\mathcal{O}(\alpha^2)\,,\qquad Q_{n}=Q_{n0}+\mathcal{O}(\alpha^2)\,,\qquad J=J_{0}+\mathcal{O}(\alpha^2)\,.
\end{equation}
We find that the identity,
\begin{equation}
    \Delta S\Big|_{M,Q_{n},J}=-\Delta I\Big|_{T,\Phi_{n},\Omega}\,,
\end{equation}
remains intact.

\section{Conclusion}

In this paper, we considered Einstein gravity in four dimensions extended with a cubic Riemann curvature invariant and studied how this term modifies perturbatively the thermodynamics of the original Ricci-flat Taub-NUT and Kerr-Taub-NUT metrics.

The perturbative solution for the Taub-NUT metric can be obtained straightforwardly owing to its $SU(2)$ isometry. However, the exact higher-order solution for the Kerr-Taub-NUT metric is unlikely to be found, even at the linear perturbative order. We thus adopted the RS method to compute the leading-order correction to the thermodynamics. The RS method does not require us to construct explicit perturbative solutions and was well established for asymptotically flat and AdS black holes. However, Taub-NUT or Kerr-Taub-NUT are not asymptotically flat nor AdS, it is therefore useful to verify the RS method using the simpler Taub-NUT solution by explicitly constructing the perturbative solution. Furthermore, there are multiple proposals of thermodynamics for Taub-NUT or Kerr-Taub-NUT metrics in literature. We adopted the proposal of \cite{Liu:2022wku} for the three reasons given in the introduction. (We also gave some discussions on a few other proposals in the appendix.) One important advantage of this proposal in \cite{Liu:2022wku} is that three quantities $(T, S)$ and $\Phi_n$, together with the Euclidean action, can be independently obtained even in the approach of finding perturbative solution to the higher-derivative curvature correction. We found that the RS method matched precisely the perturbative solution approach. This then provides a convincing validation of the thermodynamic approach \cite{Liu:2022wku}.

We then applied the RS method directly on Kerr-Taub-NUT and obtained the leading-order correction of the cubic curvature invariant to the black hole thermodynamics. The general expressions are rather complicated, but the corrections in the extremal limit reduce significantly. In future directions, one may construct the perturbed Kerr-Taub-NUT solution numerically in higher derivative gravity to verify our results and also the proposal of \cite{Liu:2022wku}. This also allows to check whether the near-horizon geometry in the extremal limit becomes irrational as in the case of \cite{Mao:2023qxq}. Furthermore, our results can be generalized to Taub-NUT AdS spacetime. In asymptotically AdS spacetime, the Reall-Santos method becomes more subtle. In \cite{Hu:2023gru} and \cite{Xiao:2023two}, two different improved RS methods are proposed, which have been successfully applied in many cases \cite{Ma:2024ynp,Wu:2024iiz,Xiao:2023two}. We can construct the perturbed Taub-NUT AdS solution and compare the results from the two methods.

\section{Acknowledge}

We are grateful to Peng-Ju Hu and Liang Ma for the useful discussion. This work is supported in part by NSFC (National Natural Science Foundation of China) Grants No.~12075166, No.~11935009,  No.~12375052 and Tianjin University Graduate Liberal Arts and Sciences Innovation Award Program (2023) No. B1-
2023-005.

\section*{Appendix}
\appendix

\section{Other proposals of Taub-NUT thermodynamics}

There have been different proposals for deciphering the thermodynamics of Taub-NUT black holes in literature. In the main text, we adopt the proposal of \cite{Liu:2022wku}, which perhaps is the only proposal satisfying all the three criteria listed in the Introduction. Nevertheless, we would like explore in this appendix how some other proposals fare under the cubic invariant perturbation. It should be remarked that the Euclidean action and its correction \eqref{RSres} based on the RS method are the same for all the proposals, but they lead to different thermodynamic quantities in different proposals, depending on the choices of the thermodynamic variables. However, using the perturbative method to verify the RS method may not be possible, if the proposal itself does not tell us how to calculate the thermodynamic variables from the perturbative solutions.

\subsection{The first version proposed in  \cite{Hennigar:2019ive}}

The first version was proposed in \cite{Hennigar:2019ive}, where the thermodynamic quantities are
\begin{equation}
    M_{0}=m\,,\qquad \Phi_{n0}=\frac{1}{8\pi n}\,,\qquad Q_{n0}=-\frac{4\pi n^3}{r_{0}}.
\end{equation}
The temperature and entropy are the same as before, since there is no controversy in these two quantities. The free energy associated with the Euclidean action is the Gibbs free energy, with $(T_0, \Phi_{n0})$ as its variables. This proposal thus satisfies one of the three criteria listed in Introduction, but it suffers from (1) the mass can be arbitrarily negative and (2) there is no smooth $n\rightarrow 0$ limit. Similar to the example we discussed in the main text, the RS method can also be established in this case. The $(T_0, \Phi_{n0})$ are fixed with the redefined constants $\tilde{r}_{0}$ and $\tilde{n}$. The modified NUT charge and mass are
\begin{equation}
Q_{n}=-\frac{4\pi \tilde{n}^3}{\tilde{r}_{0}}
+\alpha\frac{96\pi\tilde{n}^3}{7\tilde{r}_{0}(\tilde{r}_{0}^2
+\tilde{n}^2)^2}\,,\qquad M=\tilde{m}+\alpha\frac{7\tilde{r}_{0}^2-5\tilde{n}^2}
{7\tilde{r}_{0}^3(\tilde{r}_{0}^2+\tilde{n}^2)}\,.
\end{equation}
We then focus on the situation with fixed mass and NUT charge, the redefined constants are given by
\begin{equation}
    \tilde{r}_{0}\rightarrow \hat{r}_{0}+\alpha \delta \tilde{r}_{0}\,,\quad\tilde{n}\rightarrow \hat{n}+\alpha\delta \tilde{n}\,,\quad
    \delta \tilde{r}_{0}=\frac{6(5\hat{n}^4+6\hat{n}^2\hat{r}_{0}^2
    -7\hat{r}_{0}^4)}{7\hat{r}_{0}(\hat{r}_{0}^2+\hat{n}^2)^2
    (\hat{n}^2+3\hat{r}_{0}^2)}\,,\quad \delta \tilde{n}=\frac{10\hat{n}}{7\hat{r}_{0}^2(3\hat{r}_{0}^2+\hat{n}^2)}\,.
\end{equation}
Under the redefinition, we obtain
\bea
     M &=& \frac{\hat{r}_{0}^2-\hat{n}^2}{2\hat{r}_{0}}+\mathcal{O}(\alpha^2)\,,
     \qquad Q_{n}=-\frac{4\pi\hat{n}^3}{\hat{r}_{0}}+\mathcal{O}(\alpha^2)\,,\nn\\
    T &=& \frac{1}{4\pi \hat{r}_{0}}+\frac{3(-5\hat{n}^4-6 \hat{n}^2 \hat{r}_{0}^2 + 7 \hat{r}_{0}^4)}{14\pi \hat{r}_{0}^3 (\hat{n}^2 + \hat{r}_{0}^2)^2 (\hat{n}^2 + 3 \hat{r}_{0}^2)}\,,\nn\\
\Phi_{n} &=& \frac{1}{8\pi\hat{n}}-\frac{5\alpha}{28\pi\hat{n}\hat{r}_{0}^2
(\hat{n}^2+3\hat{r}_{0}^2)}\,,\qquad
S = \pi(\hat{r}_{0}^2+\hat{n}^2)+\frac{2\pi\alpha
(7\hat{r}_{0}^2-5\hat{n}^2)}{7\hat{r}_{0}^2(\hat{r}_{0}^2+\hat{n}^2)}\,.
\eea
We also verified the identity
\begin{equation}
    \Delta S\Big|_{M,Q_{n}}=\frac{2\pi\alpha(7\hat{r}_{0}^2-
    5\hat{n}^2)}{7\hat{r}_{0}^2(\hat{r}_{0}^2+\hat{n}^2)}=-\Delta I\Big|_{T,\Phi_{n}}.
\end{equation}
In the second approach, the perturbative solution and corresponding Gibbs free energy were already presented in section 3. The temperature and entropy can be determined by the standard method. We cannot make decisions for \cite{Hennigar:2019ive} how the $\Phi_n$ should be corrected; however, if we assume that it is unmodified by the perturbative solution, then the resulting perturbed thermodynamics is equivalent to the RS method.

\subsection{The Second Version Proposed in \cite{Hennigar:2019ive}}

In \cite{Hennigar:2019ive}, the authors also proposed a second version. The leading thermodynamic quantities are
\begin{equation}
M_{0}=m\,,\qquad T_{0}=\frac{1}{4\pi r_{0}}\,,\qquad S_{0}=\pi(r_{0}^2-n^2)\,,\qquad \Phi_{n0}=-\frac{n}{2}\,,\qquad Q_{n0}=\frac{n}{r_{0}}\,.
\end{equation}
Note that the entropy in this version is no longer proportional to the area of the horizon. The free energy associated with the Euclidean action is now Helmholtz type, namely
\begin{equation}\label{rm2}
    F_{0}(T_0, \Phi_{n0})=M_{0}-T_{0}S_{0}\,.
\end{equation}
Therefore, the RS method requires that the temperature and NUT charge fixed under the perturbation, and the remaining quantities can be read off as
\begin{equation}
    S_{\rm RS}=-(\frac{\partial F_{\rm RS}}{\partial T_{0}})\Big|_{Q_{n}}\,,\qquad \Phi_{n\rm RS}=(\frac{\partial F_{\rm RS}}{\partial Q_{n0}})\Big|_{T}\,,\qquad M_{\rm RS}=F_{\rm RS}+T_{0}S_{\rm RS}\,.
\end{equation}
This leads to
\bea
    S_{\rm RS} &=&\pi(\tilde{r}_{0}^2-\tilde{n}^2)+\alpha \frac{6\pi(7\tilde{r}_{0}^2-5\tilde{n}^2)}
    {7\tilde{r}_{0}^2(\tilde{r}_{0}^2+\tilde{n}^2)}\,,\quad
    \Phi_{n\rm RS}=-\frac{\tilde{n}}{2}+\alpha\frac{12}{7\tilde{n}
    (\tilde{r}_{0}^2+\tilde{n}^2)^2}\,,\cr
    M_{\rm RS}&=&\tilde{m}+\alpha\frac{7\tilde{r}_{0}^2
    -5\tilde{n}^2}{7\tilde{r}_{0}^3(\tilde{r}_{0}^2+\tilde{n}^2)}\,.
\eea
If we would like to consider fixed $(M,Q_{n})$, which requires the field redefinition
\begin{equation}
    \tilde{r}_{0}=\hat{r}_{0}+\alpha \delta \tilde{r}_{0}\,,\quad\tilde{n}= \hat{n}+\alpha\delta \tilde{n}\,,\qquad
    \delta \tilde{r}_{0}=\frac{2(5\hat{n}^2-7 \hat{r}_{0}^2)}{7\hat{r}_{0}(\hat{r}_{0}^4-\hat{n}^4)}\,,\quad\delta \tilde{n}=\frac{2\hat{n}(5\hat{n}^2-7 \hat{r}_{0}^2)}{7\hat{r}_{0}^2(\hat{r}_{0}^4-\hat{n}^4)},.
\end{equation}
We find that the following relation is satisfied
\begin{equation}
    \Delta S\Big|_{M,Q_{n}}=\frac{2\pi\alpha(7\hat{r}_{0}^2-5\hat{n}^2)}{7\hat{r}_{0}^2
    (\hat{r}_{0}^2+\hat{n}^2)}=-\Delta I\Big|_{T,Q_{n}}\,.
\end{equation}
In other words, the NUT charge can be treated literally as a constant, instead of being a thermodynamic variable.

In this thermodynamic proposal, the entropy does not satisfy the Wald formalism, making it more or less impossible to determine the correction to the entropy from the perturbative solution. With an additional ambiguity of determining the correction to $Q_n$ independently, any {\it ad hoc} method to make the results consistent with the RS method will not be likely to be convincing. We therefore are not sure whether the RS method is still applicable in this case.

\subsection{Multiple Hair Version}

Recently, a multiple hair interpretation of the Taub-NUT thermodynamics was proposed in \cite{Wu:2019pzr}. They introduce a new conserved charge $J=mn$ that is analogous to the Kerr black hole. In their prescription, the first law can be written as
\begin{equation}\label{wd}
    dm=TdS+\omega dJ+\Phi_{n}dn.
\end{equation}
with
\begin{equation}
    \omega=\frac{n}{r_{0}^2+n^2}~,~~\Phi_{n}=\frac{-2nr_{0}}{r_{0}^2+n^2}.
\end{equation}
The free energy associated with the Euclidean action then satisfies.
\begin{equation}
    F_{0}=m-T_{0}S_{0}-\omega_{0}J_{0}-\frac{n}{2}\Phi_{0}
\end{equation}
In this proposal, there are three thermodynamic variables $(T_0, \omega, \Phi_n)$ associated with the free energy, and they are not independent, but parameterized by two parameters $(r_0,n)$ only. Consequently, one cannot read off the conjugate thermodynamic quantities from the free energy, even if the reverse procedure of finding the free energy from the thermodynamic quantities can be achieved.

The key to the RS method is to find the higher-order correction to the free energy where the variables are fixed. We then read off the thermodynamic quantities conjugate to the variables from the differential relation. Thus the RS method is inapplicable to this proposal.


\begin{thebibliography}{99}

\bibitem{Taub:1950ez}
A.H.~Taub,
``Empty space-times admitting a three parameter group of motions,''
Annals Math. \textbf{53}, 472-490 (1951)
doi:10.2307/1969567

\bibitem{Newman:1963yy}
E.~Newman, L.~Tamburino and T.~Unti,
``Empty space generalization of the Schwarzschild metric,''
J. Math. Phys. \textbf{4}, 915 (1963)
doi:10.1063/1.1704018

\bibitem{Misner:1963fr}
C.W.~Misner,
``The flatter regions of Newman, Unti and Tamburino's generalized Schwarzschild space,''
J. Math. Phys. \textbf{4}, 924-938 (1963)
doi:10.1063/1.1704019

\bibitem{Clement:2015cxa}
G.~Cl\'ement, D.~Gal'tsov and M.~Guenouche,
``Rehabilitating space-times with NUTs,''
Phys. Lett. B \textbf{750}, 591-594 (2015)
doi:10.1016/j.physletb.2015.09.074
[arXiv:1508.07622 [hep-th]].

\bibitem{Clement:2015aka}
G.~Cl\'ement, D.~Gal'tsov and M.~Guenouche,
``NUT wormholes,''
Phys. Rev. D \textbf{93}, no.2, 024048 (2016)
doi:10.1103/PhysRevD.93.024048
[arXiv:1509.07854 [hep-th]].

\bibitem{Gibbons:1979xm}
G.W.~Gibbons and S.W.~Hawking,
``Classification of gravitational instanton symmetries,''
Commun. Math. Phys. \textbf{66}, 291-310 (1979)
doi:10.1007/BF01197189

\bibitem{Plebanski:1975xfb}
J.F.~Pleba\~nski,
``A class of solutions of Einstein-Maxwell equations,''
Annals Phys. \textbf{90}, no.1, 196-255 (1975)
doi:10.1016/0003-4916(75)90145-1

\bibitem{Chakraborty:2017nfu}
C.~Chakraborty and S.~Bhattacharyya, ``Does the gravitomagnetic monopole exist? A clue from a black hole x-ray binary,''
Phys. Rev. D \textbf{98}, no.4, 043021 (2018)
doi:10.1103/ PhysRevD.98.043021
[arXiv:1712.01156 [astro-ph.HE]].

\bibitem{Chakraborty:2019rna}
C.~Chakraborty and S.~Bhattacharyya, ``Circular orbits in Kerr-Taub-NUT spacetime and their implications for accreting black holes and naked singularities,''
JCAP \textbf{05}, 034 (2019)
doi:10.1088/1475-7516/2019/05/034
[arXiv:1901.04233 [astro-ph.HE]].
\bibitem{Ghasemi-Nodehi:2021ipd}
M.~Ghasemi-Nodehi, C.~Chakraborty, Q.~Yu and Y.~Lu,
``Investigating the existence of gravitomagnetic monopole in M87*,''
Eur. Phys. J. C \textbf{81}, no.10, 939 (2021)
doi:10.1140/ epjc/s10052-021-09696-3
[arXiv:2109.14903 [astro-ph.HE]].


\bibitem{Chakraborty:2022ltc}
C.~Chakraborty and S.~Bhattacharyya,
``Primordial black holes having gravitomagnetic monopole,''
Phys. Rev. D \textbf{106}, no.10, 103028 (2022)
doi:10.1103/PhysRevD.106.103028
[arXiv:2211.03610 [astro-ph.HE]].
\bibitem{Chakraborty:2023wpz}
C.~Chakraborty and B.~Mukhopadhyay,
``Geometric phase in Taub-NUT spacetime,''
Eur. Phys. J. C \textbf{83}, no.10, 937 (2023)
doi:10.1140/epjc/s10052-023-12070-0
[arXiv:2306.16318 [gr-qc]].

\bibitem{Hennigar:2019ive}
R.A.~Hennigar, D.~Kubiz\v{n}\'ak and R.B.~Mann,
``Thermodynamics of Lorentzian Taub-NUT spacetimes,''
Phys. Rev. D \textbf{100}, no.6, 064055 (2019)
doi:10.1103/PhysRevD.100.064055
[arXiv:1903.08668 [hep-th]].
\bibitem{Bordo:2019slw}
A.~B.~Bordo, F.~Gray and D.~Kubiz\v{n}\'ak,
``Thermodynamics and phase transitions of NUTty dyons,''
JHEP \textbf{07}, 119 (2019)
doi:10.1007/JHEP07(2019)119
[arXiv:1904.00030 [hep-th]].

\bibitem{BallonBordo:2019vrn}
A.~Ballon Bordo, F.~Gray, R.A.~Hennigar and D.~Kubiz\v{n}\'ak,
``The first law for rotating NUTs,''
Phys. Lett. B \textbf{798}, 134972 (2019)
doi:10.1016/j.physletb.2019.134972
[arXiv:1905.06350 [hep-th]].

\bibitem{Durka:2019ajz}
R.~Durka,
``The first law of black hole thermodynamics for Taub-NUT spacetime,''
Int. J. Mod. Phys. D \textbf{31}, no.04, 2250021 (2022)
doi:10.1142/S0218271822500213
[arXiv:1908.04238 [gr-qc]].
\bibitem{Wu:2019pzr}
S.Q.~Wu and D.~Wu,
``Thermodynamical hairs of the four-dimensional Taub-Newman-Unti-Tamburino spacetimes,''
Phys. Rev. D \textbf{100}, no.10, 101501 (2019)
doi:10.1103/Phys RevD.100.101501
[arXiv:1909.07776 [hep-th]].

\bibitem{Chen:2019uhp}
Z.~Chen and J.~Jiang,
``General Smarr relation and first law of a NUT dyonic black hole,''
Phys. Rev. D \textbf{100}, no.10, 104016 (2019)
doi:10.1103/PhysRevD.100.104016
[arXiv:1910.10107 [hep-th]].

\bibitem{BallonBordo:2020mcs}
A.~Ballon Bordo, F.~Gray and D.~Kubiz\v{n}\'ak,
``Thermodynamics of Rotating NUTty dyons,''
JHEP \textbf{05}, 084 (2020)
doi:10.1007/JHEP05(2020)084
[arXiv:2003.02268 [hep-th]].
\bibitem{Awad:2020dhy}
A.~Awad and S.~Eissa,
``Topological dyonic Taub-Bolt/NUT-AdS solutions: Thermodynamics and first law,''
Phys. Rev. D \textbf{101}, no.12, 124011 (2020)
doi:10.1103/PhysRevD. 101.124011
[arXiv:2007.10489 [gr-qc]].

\bibitem{Abbasvandi:2021nyv}
N.~Abbasvandi, M.~Tavakoli and R.B.~Mann,
``Thermodynamics of dyonic NUT charged black holes with entropy as Noether charge,''
JHEP \textbf{08}, 152 (2021)
doi:10.1007/ JHEP08(2021)152
[arXiv:2107.00182 [hep-th]].

\bibitem{Frodden:2021ces}
E.~Frodden and D.~Hidalgo,
``The first law for the Kerr-NUT spacetime,''
Phys. Lett. B \textbf{832}, 137264 (2022)
doi:10.1016/j.physletb.2022.137264
[arXiv:2109.07715 [hep-th]].
\bibitem{Wu:2022rmx}
D.~Wu and S.Q.~Wu,
``Consistent mass formulas for the four-dimensional dyonic NUT-charged spacetimes,''
Phys. Rev. D \textbf{105}, no.12, 124013 (2022)
doi:10.1103/PhysRevD. 105.124013
[arXiv:2202.09251 [gr-qc]].

\bibitem{Godazgar:2022jxm}
M.~Godazgar and S.~Guisset,
``Dual charges for AdS spacetimes and the first law of black hole mechanics,''
Phys. Rev. D \textbf{106}, no.2, 024022 (2022)
doi:10.1103/PhysRevD. 106.024022
[arXiv:2205.10043 [hep-th]].

\bibitem{Awad:2022jgn}
A.~Awad and S.~Eissa,
``Lorentzian Taub-NUT spacetimes: Misner string charges and the first law,''
Phys. Rev. D \textbf{105}, no.12, 124034 (2022)
doi:10.1103/PhysRevD.105.124034
[arXiv:2206.09124 [hep-th]].
\bibitem{Wu:2022mlz}
D.~Wu and S.Q.~Wu,
``Consistent mass formulas for higher even-dimensional Taub-NUT spacetimes and their AdS counterparts,''
Phys. Rev. D \textbf{108}, no.6, 064034 (2023)
doi:10.1103/PhysRevD.108.064034
[arXiv:2209.01757 [hep-th]].
\bibitem{Wu:2022xpp}
D.~Wu and S.Q.~Wu,
``Revisiting mass formulas of the four-dimensional Reissner-Nordstr\"om-NUT-AdS solutions in a different metric form,''
Phys. Lett. B \textbf{846}, 138227 (2023)
doi:10.1016/j.physletb.2023.138227
[arXiv:2210.17504 [gr-qc]].
\bibitem{Wu:2023woq}
S.Q.~Wu and D.~Wu,
``Consistent mass formulas for higher even-dimensional Reissner-Nordstr\"om-NUT-AdS spacetimes,''
Phys. Rev. D \textbf{108}, no.6, 064035 (2023)
doi:10.1103/ PhysRevD.108.064035
[arXiv:2306.00062 [gr-qc]].

\bibitem{Liu:2022wku}
H.S.~Liu, H.~L\"u and L.~Ma,
``Thermodynamics of Taub-NUT and Plebanski solutions,''
JHEP \textbf{10}, 174 (2022)
doi:10.1007/JHEP10(2022)174
[arXiv:2208.05494 [gr-qc]].
\bibitem{Liu:2023uqf}
J.F.~Liu and H.S.~Liu,
``Thermodynamics of Taub-NUT-AdS spacetimes,''
Eur. Phys. J. C \textbf{84}, no.5, 515 (2024)
doi:10.1140/epjc/s10052-024-12826-2
[arXiv:2309.01609 [hep-th]].



\bibitem{Jiang:2019yzs}
J.~Jiang, B.~Deng and X.W.~Li,
``Holographic complexity of charged Taub-NUT-AdS black holes,''
Phys. Rev. D \textbf{100}, no.6, 066007 (2019)
doi:10.1103/PhysRevD.100.066007
[arXiv:1908.06565 [hep-th]].

\bibitem{Chen:2023eio}
S.~Chen, Y.~Pei, L.~Li and T.~Yang,
``Charged Taub-NUT-AdS Black Holes in $f(R)$ Gravity and Holographic Complexity,''
Int. J. Theor. Phys. \textbf{62}, no.2, 16 (2023)
doi:10.1007/s10773-023-05280-5
\bibitem{Perry:2022udk}
M.J.~Perry and M.J.~Rodriguez,
``CFT duals of Kerr-Taub-NUT and beyond,''
[arXiv: 2205.09146 [hep-th]].
\bibitem{Siahaan:2022jrl}
H.M.~Siahaan,
``Hidden conformal symmetry and pair production near the cosmological horizon in Kerr-Newman-Taub-NUT-de Sitter spacetime*,''
Chin. Phys. C \textbf{47}, no.3, 035104 (2023)
doi:10.1088/1674-1137/aca95d
[arXiv:2212.03058 [gr-qc]].


\bibitem{Yang:2023hll}
S.J.~Yang, W.D.~Guo, S.W.~Wei and Y.X.~Liu,
``First law of black hole thermodynamics and the weak cosmic censorship conjecture for Kerr-Newman Taub-NUT black holes,''
Eur. Phys. J. C \textbf{83}, no.12, 1111 (2023)
doi:10.1140/epjc/s10052-023-12265-5
[arXiv:2306.05266 [gr-qc]].

\bibitem{Wu:2024ucf}
P.Y.~Wu, H.~Khodabakhshi and H.~L\"u,
``Weak cosmic censorship conjecture cannot be violated in Gedanken experiments,''
[arXiv:2408.09444 [gr-qc]].

\bibitem{tHooft:1974toh}
G.~'t Hooft and M.J.~G.~Veltman,
``One loop divergencies in the theory of gravitation,''
Ann. Inst. H. Poincare Phys. Theor. A \textbf{20}, 69-94 (1974)

\bibitem{Stelle:1976gc}
K.S.~Stelle,
``Renormalization of higher derivative quantum gravity,''
Phys. Rev. D \textbf{16}, 953-969 (1977)
doi:10.1103/PhysRevD.16.953

\bibitem{Lu:2015cqa}
H.~L\"u, A.~Perkins, C.N.~Pope and K.S.~Stelle,
``Black holes in higher-derivative gravity,''
Phys. Rev. Lett. \textbf{114}, no.17, 171601 (2015)
doi:10.1103/PhysRevLett.114.171601
[arXiv:1502.01028 [hep-th]].

\bibitem{Bueno:2018uoy}
P.~Bueno, P.A.~Cano, R.A.~Hennigar and R.~B.~Mann,
``NUTs and bolts beyond Lovelock,''
JHEP \textbf{10}, 095 (2018)
doi:10.1007/JHEP10(2018)095
[arXiv:1808.01671 [hep-th]].
\bibitem{Brihaye:2018bgc}
Y.~Brihaye, C.~Herdeiro and E.~Radu,
``The scalarised Schwarzschild-NUT spacetime,''
Phys. Lett. B \textbf{788}, 295-301 (2019)
doi:10.1016/j.physletb.2018.11.022
[arXiv:1810.09560 [gr-qc]].
\bibitem{Ibadov:2021oqf}
R.~Ibadov, B.~Kleihaus, J.~Kunz and S.~Murodov,
``Wormhole solutions with NUT charge in higher curvature theories,''
Arab. J. Math. \textbf{11}, no.1, 31-41 (2022)
doi:10.1007/s40065-021-00350-0
[arXiv:2111.09628 [gr-qc]].
\bibitem{Butler:2023tyt}
M.~Butler and M.~Ghezelbash,
``NUT solutions in Einstein-Maxwell-scalar-Gauss-Bonnet gravity,''
Phys. Rev. D \textbf{109}, no.4, 044018 (2024)
doi:10.1103/PhysRevD.109.044018
[arXiv:2310.04568 [gr-qc]].
\bibitem{Mukherjee:2021erg}
S.~Mukherjee and N.~Dadhich,
``Pure Gauss\textendash{}Bonnet NUT black hole solution: I,''
Eur. Phys. J. C \textbf{82}, no.4, 302 (2022)
doi:10.1140/epjc/s10052-022-10256-6
[arXiv:2101.02958 [gr-qc]].
\bibitem{Mukherjee:2020lld}
S.~Mukherjee and N.~Dadhich,
``Pure Gauss\textendash{}Bonnet NUT black hole with and without non-central singularity,''
Eur. Phys. J. C \textbf{81}, no.5, 458 (2021)
doi:10.1140/epjc/s10052-021-09242-1
[arXiv:2012.15560 [gr-qc]].
\bibitem{Chen:2024hsh}
Y.Q.~Chen and H.S.~Liu,
``New Taub-NUT black holes with massive Spin-2 hair,''
[arXiv:2406.03692 [gr-qc]].

\bibitem{Reall:2019sah}
H.S.~Reall and J.E.~Santos,
``Higher derivative corrections to Kerr black hole thermodynamics,''
JHEP \textbf{04}, 021 (2019)
doi:10.1007/JHEP04(2019)021
[arXiv:1901.11535 [hep-th]].


\bibitem{Xiao:2022auy}
Y.~Xiao,
``First order corrections to the black hole thermodynamics in higher curvature theories of gravity,''
Phys. Rev. D \textbf{106}, no.6, 064041 (2022)
doi:10.1103/ PhysRevD.106.064041
[arXiv:2207.00967 [gr-qc]].
\bibitem{Ma:2023qqj}
L.~Ma, Y.~Pang and H.~L\"u,
``Higher derivative contributions to black hole thermodynamics at NNLO,''
JHEP \textbf{06}, 087 (2023)
doi:10.1007/JHEP06(2023)087
[arXiv:2304.08527 [hep-th]].


\bibitem{Endlich:2017tqa}
S.~Endlich, V.~Gorbenko, J.~Huang and L.~Senatore,
``An effective formalism for testing extensions to General Relativity with gravitational waves,''
JHEP \textbf{09}, 122 (2017)
doi:10.1007/JHEP09(2017)122
[arXiv:1704.01590 [gr-qc]].
\bibitem{Cano:2019ore}
P.A.~Cano and A.~Ruip\'erez,
``Leading higher-derivative corrections to Kerr geometry,''
JHEP \textbf{05}, 189 (2019)
[erratum: JHEP \textbf{03}, 187 (2020)]
doi:10.1007/JHEP05(2019)189
[arXiv:1901.01315 [gr-qc]].

\bibitem{Wald:1993nt}
R.M.~Wald,
``Black hole entropy is the Noether charge,''
Phys. Rev. D \textbf{48}, no.8, R3427-R3431 (1993)
doi:10.1103/PhysRevD.48.R3427
[arXiv:gr-qc/9307038 [gr-qc]].
\bibitem{Iyer:1994ys}
V.~Iyer and R.M.~Wald,
``Some properties of Noether charge and a proposal for dynamical black hole entropy,''
Phys. Rev. D \textbf{50}, 846-864 (1994)
doi:10.1103/PhysRevD.50.846
[arXiv:gr-qc/9403028 [gr-qc]].

\bibitem{Mao:2023qxq}
Q.Y.~Mao, L.~Ma and H.~L\"u,
``Horizon as a natural boundary,''
Phys. Rev. D \textbf{109}, no.8, 084053 (2024)
doi:10.1103/PhysRevD.109.084053
[arXiv:2307.14458 [hep-th]].

\bibitem{Hu:2023gru}
P.J.~Hu, L.~Ma, H.~L\"u and Y.~Pang,
``Improved Reall-Santos method for AdS black holes in general 4-derivative gravities,''
[arXiv:2312.11610 [hep-th]].
\bibitem{Xiao:2023two}
Y.~Xiao and Y.Y.~Liu,
``First order corrections to black hole thermodynamics: a simple approach enhanced,''
[arXiv:2312.07127 [gr-qc]].


\bibitem{Ma:2024ynp}
L.~Ma, P.J.~Hu, Y.~Pang and H.~L\"u,
``The surprising effectiveness of Weyl gravity in probing quantum corrections to AdS black holes,''
[arXiv:2403.12131 [hep-th]].

\bibitem{Wu:2024iiz}
P.Y.~Wu and H.~L\"u,
``Quadratic curvature correction and its breakdown to thermodynamics of rotating black holes,''
[arXiv:2405.04576 [hep-th]].
\end{thebibliography}
\end{document}